\documentclass[conference]{IEEEtran}
\IEEEoverridecommandlockouts
\usepackage{cite}
\usepackage{amsmath,amssymb,amsfonts}
\usepackage{algorithmic}
\usepackage{graphicx}
\usepackage{textcomp}
\usepackage{xcolor}
\usepackage[hyphens]{url}
\usepackage{comment}
\usepackage[breaklinks=true]{hyperref}
\usepackage{booktabs}
\usepackage{tabularx}
\usepackage{graphicx}
\usepackage{colortbl}
\usepackage[table]{xcolor}
\usepackage{rotating}
\usepackage{array}  
\usepackage{amssymb} 
\usepackage{graphicx} 

\definecolor{lightgray}{gray}{0.9}

\newcolumntype{Z}{>{\hsize=1.3\hsize\columncolor{lightgray}}X}

\def\BibTeX{{\rm B\kern-.05em{\sc i\kern-.025em b}\kern-.08em
    T\kern-.1667em\lower.7ex\hbox{E}\kern-.125emX}}
\begin{document}

\title{CRACI: A Cloud-Native Reference Architecture for the Industrial Compute Continuum}

\author{\IEEEauthorblockN{Hai Dinh-Tuan}
\IEEEauthorblockA{\textit{Service-Centric Networking} \\
\textit{Technische Universit\"{a}t Berlin}\\
Berlin, Germany \\
hai.dinhtuan@tu-berlin.de}
}
\maketitle

\begin{abstract}
The convergence of Information Technology (IT) and Operational Technology (OT) in Industry 4.0 exposes the limitations of traditional, hierarchical architectures like ISA-95 and RAMI 4.0. Their inherent rigidity, data silos, and lack of support for cloud-native technologies impair the development of scalable and interoperable industrial systems. This paper addresses this issue by introducing CRACI, a Cloud-native Reference Architecture for the Industrial Compute Continuum. Among other features, CRACI promotes a decoupled and event-driven model to enable flexible, non-hierarchical data flows across the continuum. It embeds cross-cutting concerns as foundational pillars: Trust, Governance \& Policy, Observability, and Lifecycle Management, ensuring quality attributes are core to the design. The proposed architecture is validated through a two-fold approach: (1) a comparative theoretical analysis against established standards, operational models, and academic proposals; and (2) a quantitative evaluation based on performance data from previously published real-world smart manufacturing implementations. The results demonstrate that CRACI provides a viable, state-of-the-art architecture that utilizes the compute continuum to overcome the structural limitations of legacy models and enable scalable, modern industrial systems.

\end{abstract}


\section{Introduction}

The \textit{Fourth Industrial Revolution (Industry 4.0)} is driving an unprecedented convergence of physical production systems and digital intelligence. Modern industrial environments demand architectures that support scalable, low-latency analytics, autonomous control, and seamless integration across the entire \textit{Compute Continuum} (CC), from edge devices on the factory floor to centralized cloud platforms. However, the models that current industrial systems rely on have proven inadequate for fully harnessing the benefits of this paradigm.

Established standards such as \textit{ISA-95} and the \textit{Reference Architecture Model for Industry 4.0 (RAMI 4.0)}, while valuable for structuring enterprise-control systems, are derived from a rigid, hierarchical view. Their top-down structure creates data silos, limits real-time responsiveness, and fails to natively support the decentralized, event-driven, and microservice-based patterns that define modern cloud-native systems. This fundamental mismatch creates a significant architectural gap, resulting in inflexible, monolithic systems that are difficult to maintain, operate, and scale. While more recent industry-driven operational models like FIWARE and NOA partly address aspects of this problem, they often remain either too abstract, too technologically prescriptive, or incomplete, lacking a unified operational model for managing distributed services across heterogeneous environments. Similarly, academic proposals often present conceptual models that, while insightful, often lack the practical implementation details and operational guidance necessary for real-world adoption.

To address this critical gap, this work introduces CRACI: a Cloud-native Reference Architecture for the Compute Continuum in Industry 4.0. Instead of a rigid pyramid, CRACI proposes a decoupled, service-oriented system built on two core patterns: a \textit{hub-and-spoke model} centered on a Digital Twin repository and an \textit{event-driven communication} backbone. This design overcomes data silos and enables flexible, non-hierarchical interactions between components. A key novelty of the proposed architecture is the formal integration of four \textit{Foundational Pillars}: Trust, Governance \& Policy, Observability, and Lifecycle Management, which embed critical quality attributes directly into the system design. This approach ensures the architecture is not only aligned with cloud-native best practices but also maintains a balance between high-level abstraction (for general applicability) and detailed guidance (for concrete implementation). The primary contributions of this work are therefore threefold:

\begin{itemize}
    \item An analytical review of Industry 4.0 architectural standards and models, highlighting critical gaps in operational and governance concerns.
    
    \item The design process of CRACI, resulting in a novel cloud-native reference architecture that resolves the identified gaps through a data-centric design and the formalization of four Foundational Pillars.

    \item A qualitative and quantitative evaluation demonstrating how CRACI's design systematically addresses the limitations of existing models like ISA-95 or RAMI 4.0.
\end{itemize}

The remainder of this work is structured as follows: Section II reviews related work in industrial reference architectures. Section III presents the design methodology while Section IV introduces the CRACI architecture in detail. Section V provides the theoretical comparative analysis, followed by the quantitative evaluation in Section VI. Finally, Section VII concludes the work by summarizing the contributions and outlining directions for future work.

\section{Related Work}

To position our contribution, we review three classes of existing work, which are formal standards, industry-driven frameworks, and academic architectures to identify the specific gaps that motivate the design of CRACI.

\subsection{Formal Industry Standards}

The ISA-95 model \cite{scholten2007road} has long served as the foundational model for integrating enterprise and control systems. Based on the \textit{Purdue Enterprise Reference Architecture}, it defines a five-level automation pyramid that organizes functions from the physical process level (Level 0) up to business logistics (Level 4). While ISA-95 provides essential structural clarity, its rigid, top-down hierarchy is fundamentally ill-suited for modern, event-driven environments. It was designed before the era of cloud computing, leading to the creation of data silos, which severely impede the kind of flexible, cross-layer data access required for advanced analytics and distributed intelligence in Industry 4.0.

\begin{figure}[htbp]
    \centering
    \includegraphics[width=\columnwidth]{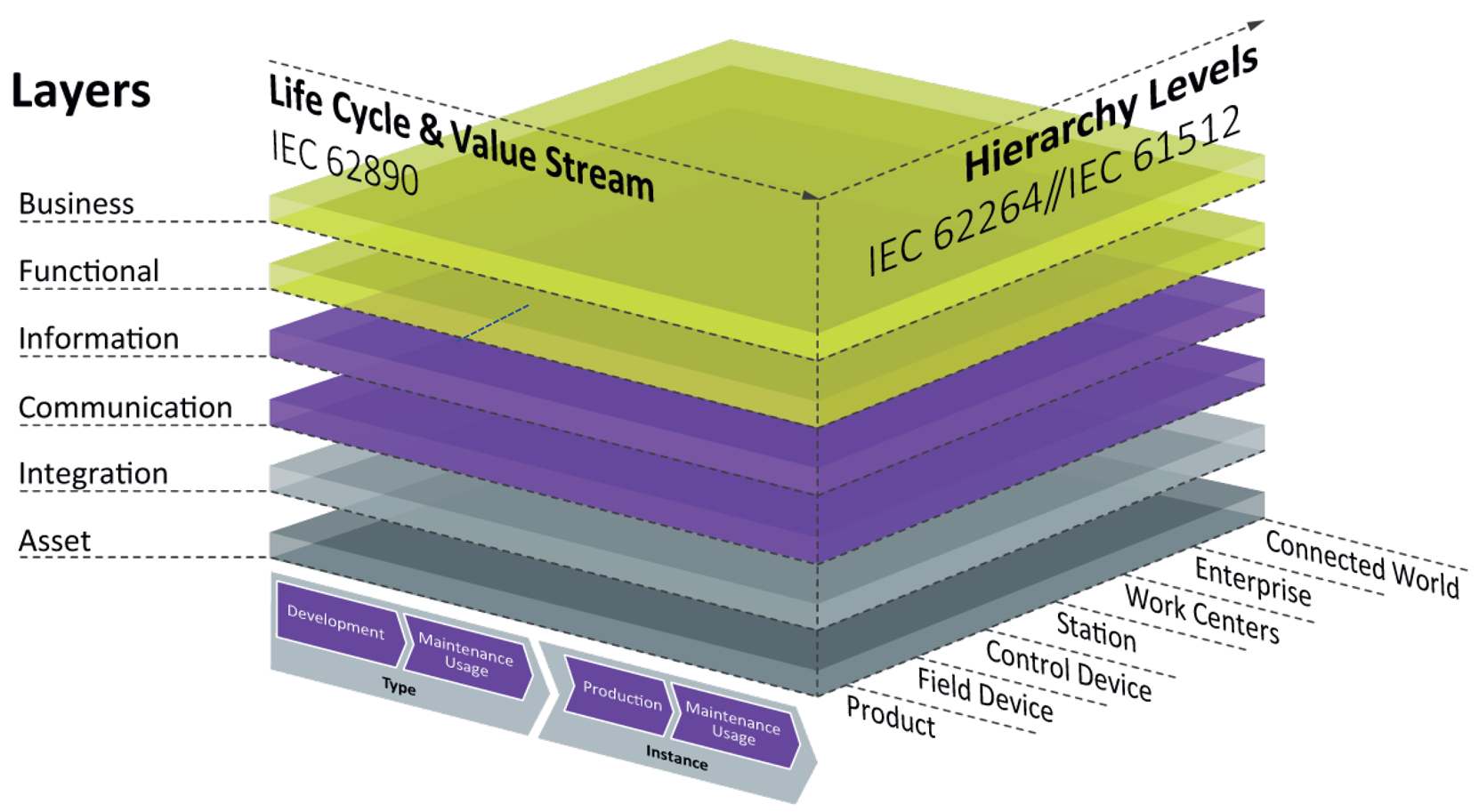}
    \caption{Reference Architecture Model Industrie 4.0 (RAMI 4.0) \cite{hankel2015reference}.}
    \label{fig:rami4}
\end{figure}

As a direct response to the needs of Industry 4.0, Germany's \textit{Plattform Industrie 4.0} introduced the \textit{Reference Architecture Model for Industry 4.0 (RAMI 4.0)} \cite{hankel2015reference}. RAMI 4.0 extends ISA-95 into a three-dimensional map as illustrated in Fig.~\ref{fig:rami4}, with six \textit{Hierarchy Levels} (based on IEC 62264~\cite{iec62264} and IEC 61512~\cite{iec61512}), incorporating a \textit{Life Cycle \& Value Stream} axis (based on IEC 62890 \cite{iec62890}) and an \textit{Architecture Layers} axis to classify components. While RAMI 4.0 provides a valuable conceptual map for structuring assets and functions, it remains a highly abstract rather than operational. It offers little guidance on how to deploy, manage, or orchestrate distributed services, and it retains the hierarchical assumptions of ISA-95, which can be incompatible with direct cross-level interactions along the value chain found in modern manufacturing practices. Furthermore, critical cross-cutting concerns like governance and trust are not explicitly embedded within the model.

\subsection{Industry-Driven Operational Frameworks}

\begin{figure}[htbp]
    \hspace{0.04\linewidth} 
    \includegraphics[width=\columnwidth]{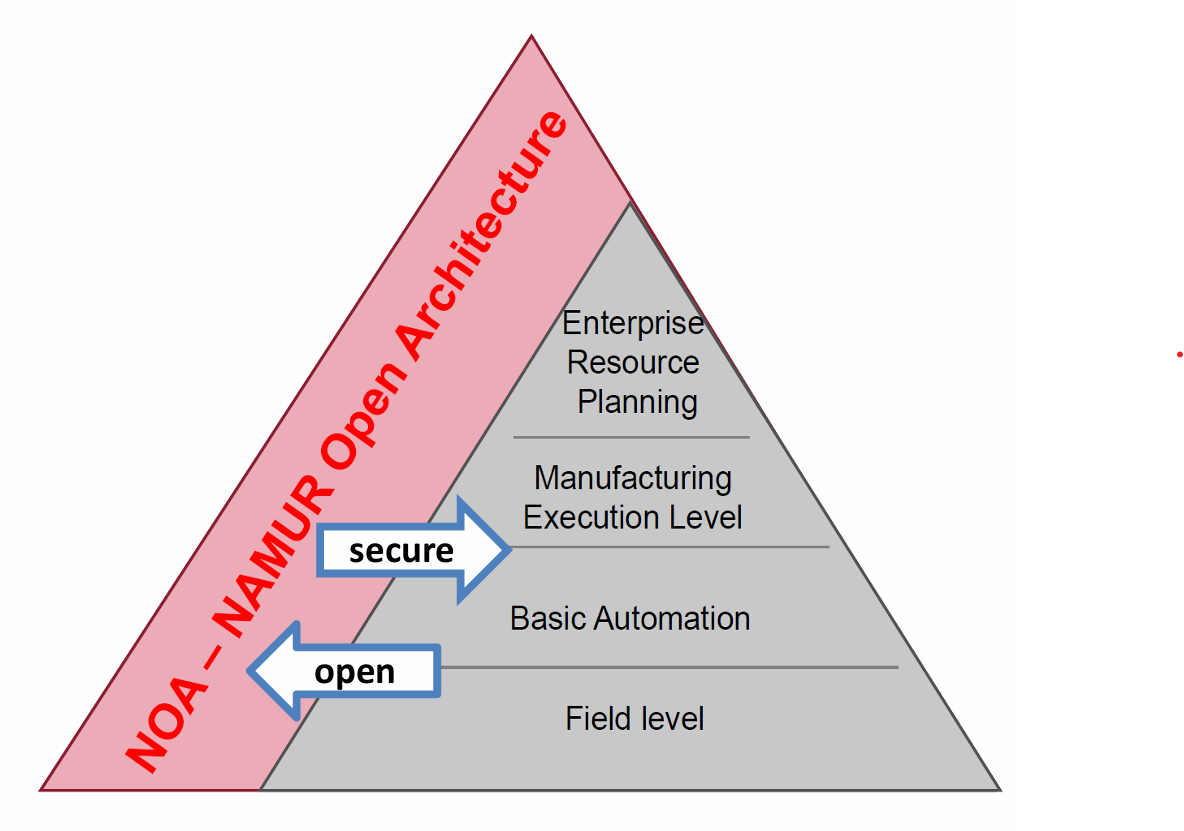}
    \caption{NAMUR Open Architecture \cite{de2020namur, klettner2017namur}.}
    \label{fig:namur}
\end{figure}

Several operational frameworks address these abstract limitations. For instance, the NAMUR Open Architecture (NOA) \cite{de2020namur, klettner2017namur} offers an approach for brownfield environments by introducing a \textit{Monitoring and Optimization} (M+O) domain that securely extracts data from the core process control domain as illustrated in Fig.~\ref{fig:namur}. However, NOA is intended as a transitional pattern, not a complete architecture. Moreover, due to its strong industrial origins, it treats the M+O domain merely as a conceptual block, lacking a concrete architectural model for the actual IT services and components that should operate within this domain.

The FIWARE framework \cite{alonso2018industrial} offers a mature, data-centric platform centered around its NGSI-LD (Next Generation Service Interface-Linked Data) Context Broker. Despite its comprehensive design, its benefits are tied to a specific ecosystem of \textit{enablers}, such as Orion Context Broker\footnote{\url{https://github.com/telefonicaid/fiware-orion}}, Cygnus\footnote{\url{https://github.com/telefonicaid/fiware-cygnus}}, Keyrock\footnote{\url{https://keyrock-fiware.github.io/}}, which creates a steep learning curve and risks technology lock-in. Furthermore, while critical concerns like trust and lifecycle management are acknowledged, they are handled externally instead of being embedded as core architectural principles.

Lastly, the AIOTI Functional Model \cite{open_continuum_eucloudedgeiot_task_force_2024_11656674}, proposed by the EUCloudEdgeIoT.eu initiative, represents a significant effort to define key capabilities for the CC, organizing them into nine functional domains like data management, orchestration, and security. Despite outlining needed capabilities, the model stays mostly conceptual and provides limited technical guidance, lacking concrete implementation patterns.

\subsection{Academic Reference Architectures}

Academic researchers have also proposed alternative architectures for Industry 4.0. For instance, Resman et al.~\cite{resman2019new} propose the LASFA model as a more user-oriented and pragmatic alternative to the complex RAMI 4.0. The LASFA architecture addresses RAMI 4.0's shortcomings by introducing a simplified, two-dimensional representation of digital twins, agents, and data flows.

The \textit{Stuttgart IT Architecture for Manufacturing (SITAM)} \cite{kassner2016stuttgart} also addresses the limitations of the hierarchical information pyramid, which include rigid integrations, siloed data, and limited shop-floor accessibility, by introducing the concept of a \textit{data-driven factory} that is agile and human-centric. As a multi-layered, service-oriented architecture, SITAM bridges the gap between abstract reference models (e.g., RAMI 4.0) and concrete implementations, providing both a full reference architecture and concrete technological recommendations.

Jiang's 8C architecture~\cite{jiang2018improved} extends the 5C architecture~\cite{lee2015cyber} by adding \textit{Coalition}, \textit{Customer}, and \textit{Content} to support horizontal integration and external collaboration. This broader scope incorporates supply chains, end customers, and lifecycle data, making it better suited to mass customization and interconnected industrial ecosystems.

\subsection{Summary and Motivation for CRACI}

The reviewed models and architectures reveal a critical gap that motivates our work. Formal standards are overly hierarchical and abstract, while industry frameworks are either functionally incomplete (NOA), technologically prescriptive (FIWARE), or lacking operational guidance (AIOTI). Academic proposals, though insightful, often emphasize functional layouts without embedding key quality attributes as design drivers. To address this, CRACI provides a flexible, technology-agnostic reference architecture that combines structural clarity with cloud-native principles, grounded in the pillars of Trust, Governance \& Policy, Observability, and Lifecycle Management.

\section{Design Methodology and Architectural Drivers}
The design of CRACI follows the \textit{Attribute-Driven Design} (ADD 3.0) methodology, which emphasizes \textit{Quality Attribute Requirements} (QARs) during the design phase~\cite{cervantes2024designing}. This approach ensures that the final design is directly driven by the core challenges and goals of the system. The key architectural drivers, derived from the gaps identified previously in our related work analysis, are outlined below.

\subsection{Quality and Functional Goals}
The primary purpose of CRACI is to create a unified and extensible reference architecture that resolves the structural limitations of traditional models (e.g., ISA-95's hierarchy) through a decoupled, event-driven design, following cloud-native principles. Critically, it aims to ensure operational robustness by treating key quality attributes as foundational design principles. These goals are refined into the following specific requirements:

\subsubsection{Quality Attribute Requirements (QARs)} These define the critical non-functional properties that directly influence the selection of architectural patterns and are realized as CRACI's Foundational Pillars.
\begin{itemize}
    \item \textbf{QAR-1 (Trustworthiness):} The system must enforce identity and data usage policies across all interactions in a distributed, multi-stakeholder environment.
    \item \textbf{QAR-2 (Governance):} The system must support data sovereignty and comply with legal and business regulations through a managed, auditable policy framework.
    \item \textbf{QAR-3 (Observability):} The system's operational state must be fully transparent through high-quality telemetry for effective monitoring and root-cause analysis.
    \item \textbf{QAR-4 (Maintainability):} The architecture must support agile and reliable evolution through automated, version-controlled lifecycle management processes.
\end{itemize}

\subsubsection{Primary Functional Requirements (FRs)} These define the specific features the system must deliver.
\begin{itemize}
    \item \textbf{FR-1 (Unified Asset Representation):} Provide a stateful Digital Twin for every asset, serving as the single source of truth and resolving data silos.
    \item \textbf{FR-2 (Edge Intelligence):} Support the deployment of low-latency analytics (e.g., condition monitoring, anomaly detection) as close to physical assets as possible.
    \item \textbf{FR-3 (Cloud-based Strategic Intelligence):} Offer cloud services for long-term, high-latency analytics such as predictive maintenance and enterprise-wide planning.
    \item \textbf{FR-4 (Business Process Orchestration):} Coordinate complex, long-running business processes (e.g., order-to-execution) through a dedicated orchestration component.
\end{itemize}

\subsection{Architectural Concerns and Constraints}
The design is bound by several key constraints (CON). The architecture must:
\begin{itemize}
    \item \textbf{CON-1:} Support hybrid edge-cloud environments, including legacy brownfield assets.
    \item \textbf{CON-2:} Be implemented with cloud-native technologies and open standards (e.g., OPC UA (Open Platform Communications Unified Architecture), MQTT (Message Queuing Telemetry Transport)) to avoid vendor lock-in.
    \item \textbf{CON-3:} Remain conceptually aligned with established frameworks like RAMI 4.0, IEC 62890, and data space principles (e.g., Gaia-X \cite{braud2021road}, Catena-X \cite{mugge2023empowering}).
\end{itemize}

Within these boundaries, the design addresses four primary architectural concerns (AC):
\begin{itemize}
    \item \textbf{AC-1 (Decomposition \& Decoupling):} To avoid ISA-95's rigid hierarchy, the architecture uses a \textit{hub-and-spoke model} for its data core and a \textit{publish/subscribe pattern} for communication.
    \item \textbf{AC-2 (Semantic Interoperability):} To handle heterogeneity, it employs a two-part strategy of \textit{protocol adaptation at the edge} and \textit{centralized semantic enrichment}.
    \item \textbf{AC-3 (Secure OT/IT Convergence):} To safely extract data from protected OT networks, it implements the secure gateway concept inspired by NOA's principles.
    \item \textbf{AC-4 (Operational Completeness):} To go beyond abstract diagrams (like in RAMI 4.0), it considers the QARs as the four \textit{Foundational Pillars} that govern every component in the system.
\end{itemize}

\section{The CRACI Reference Architecture}

\begin{figure*}[htbp]
    \centering
    \includegraphics[width=0.8\linewidth]{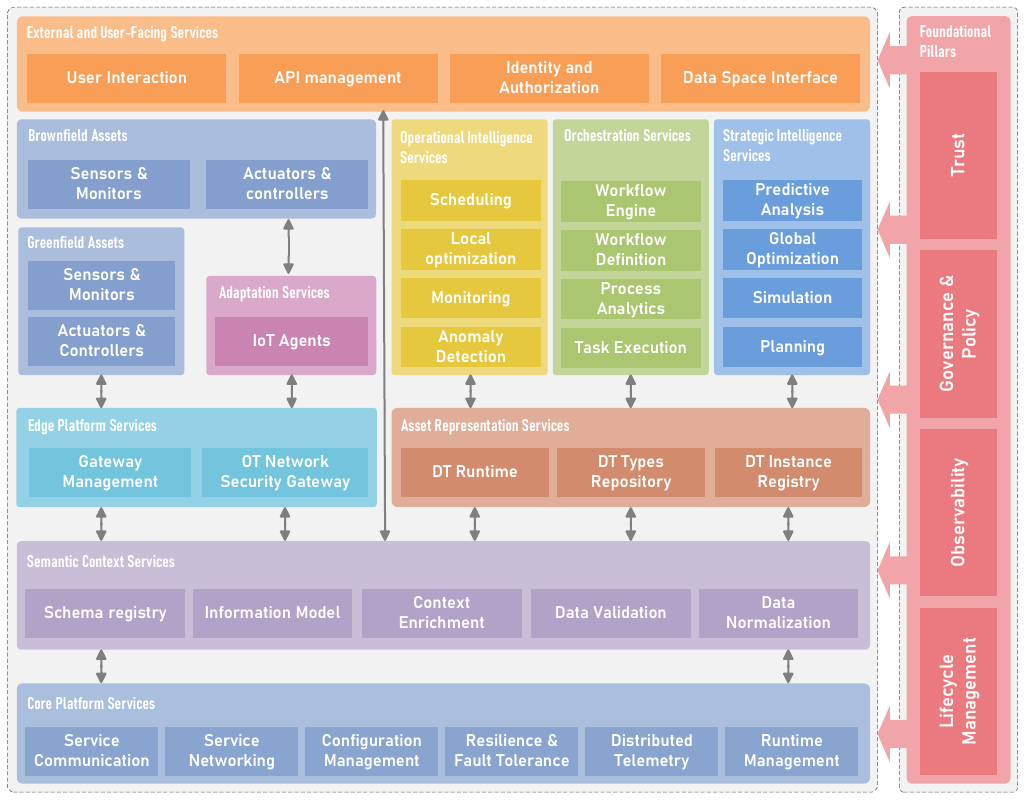}
    \caption{Cloud-native Reference Architecture for the Compute Continuum in Industry 4.0 (CRACI)}
    \label{fig:reference-architecture}
\end{figure*}

The Cloud-native Reference Architecture for the Compute Continuum in Industry 4.0 (CRACI), shown in Fig.~\ref{fig:reference-architecture}, is a blueprint designed to satisfy the architectural drivers identified previously. It is structured as a service-oriented, data-centric system that intentionally moves away from the rigid hierarchies of traditional models.

\subsection{Architectural Overview and Foundational Pillars}
The primary organizing principle is its horizontal axis, representing the physical or logical deployment location of each component. From left to right, the architecture progresses from distributed data sources to centralized cloud resources:

\begin{itemize}
    \item \textit{Shop Floor (Far Left)}: Contains physical assets and platform services (e.g., Adaptation Services, Edge Platform Services). Operational Intelligence Services can also be deployed here to meet low-latency requirements for real-time monitoring and control.
    
    \item \textit{Regional Core (Middle)}: Hosts services centralized for consistency, such as Asset Representation Services (Digital Twin Hub), and Orchestration Services.
    
    \item \textit{Cloud (Far Right)}: Contains compute-intensive, non-latency-sensitive services like Strategic Intelligence Services, model training, and global optimization.
\end{itemize}

All external interactions are managed through the \textit{External and User-Facing Services}, which extend all the way from the physical device level to central cloud, suggesting that external integration can happen anywhere across the continuum. The system is built on domain-agnostic \textit{Core Platform Services} and governed by four Foundational Pillars: \textit{Trust}, \textit{Governance \& Policy}, \textit{Observability}, and \textit{Lifecycle Management}. As direct realizations of the QARs, these pillars are embedded as system-wide principles, ensuring consistent quality attributes across all components and deployment locations:
\begin{itemize}
    \item \textit{Trust - QAR-1:} Ensures secure identity, authentication, and authorization for all interactions, forming the basis for participation in federated ecosystems like Gaia-X. In highly secured contexts, it enables the realization of Zero Trust Architecture, while remaining compatible with both conventional security frameworks (e.g., PKI- or OAuth2-based models) and emerging approaches such as Decentralized Digital Identity \cite{10597085, 10494437}.
    \item \textit{Governance \& Policy - QAR-2:} Enforces data sovereignty, regulatory compliance (e.g., GDPR (General Data Protection Regulation)), and auditable data usage policies across all distributed services.
    \item \textit{Observability - QAR-3:} Provides end-to-end system transparency by collecting and analyzing telemetry (logs, metrics, traces) from all components for root-cause analysis.
    \item \textit{Lifecycle Management - QAR-4:} Implements modern DevOps/MLOps practices (e.g., IaC (Infrastructure as Code), CI/CD (Continuous Integration and Continuous Delivery/Deployment), GitOps) to govern the automated, version-controlled lifecycle of all system artifacts, ensuring agility and reproducibility.
\end{itemize}

\subsection{Detailed Service Groups}
The architecture's service groups follow a logical data value chain, transforming raw sensor data into actionable intelligence.

\subsubsection{Data Ingestion and Edge Processing}
At the far left, \textit{Physical Assets} (both legacy Brownfield and modern Greenfield) are the origin of all data. To handle heterogeneity (\textbf{AC-2}), \textit{Adaptation Services} act as a translation layer, converting proprietary protocols (e.g., Modbus RTU/TCP, PROFIBUS, DeviceNet, etc.) into standard formats used by other services in the continuum. These data streams enter the platform through the \textit{Edge Platform Services}, which include:
\begin{itemize}
    \item an \textit{OT Network Security Gateway} that enforces a secure OT/IT boundary (\textbf{AC-3}), and
    \item a \textit{Gateway Management} service for orchestrating edge nodes.
\end{itemize}
This layer also hosts \textit{Operational Intelligence Services}, which perform low-latency analytics such as condition monitoring and anomaly detection (\textbf{FR-2}).

\subsubsection{Semantic Enrichment and Data Centering}
Once ingested, data streams flow to the \textit{Semantic Context Services}. This component is critical for achieving interoperability, as it validates, normalizes, and enriches the data with business context using a central \textit{Information Model} (a semantic graph of all assets). The now contextually rich data is consumed by the \textit{Asset Representation Services}, which function as a stateful \textit{Digital Twin Hub} (\textbf{FR-1}). This component, implementing the IEC 62890 Type/Instance model \cite{iec62890}, acts as the single source of truth, effectively breaking down the data silos inherent in hierarchical models (\textbf{AC-1}). It decouples data providers from consumers, allowing any authorized service to access a consistent, live, and updated view of the system's state.

\subsubsection{Cloud Analytics and Orchestration}
The Digital Twin Hub serves as the primary data source for higher-level services. \textit{Strategic Intelligence Services}, typically deployed in the cloud, consume historical and real-time data from the hub to perform long-term analytics like predictive maintenance and global optimization \cite{10577743} (\textbf{FR-3}). Concurrently, the \textit{Orchestration Services} component coordinates complex, stateful business processes by executing workflow models (e.g., in BPMN (Business Process Model and Notation)), or service migration \cite{10577743, 10942720}, translating high-level business events into sequences of actions across the distributed system (\textbf{FR-4}).

\subsubsection{External Interfaces}
All interactions with external actors are managed by the \textit{External and User-Facing Services}. This includes:
\begin{itemize}
    \item a \textit{Unified API Management Gateway} for programmatic access,
    \item \textit{User Interaction Services} for dashboards, and
    \item a specialized \textit{Data Space Interface} that enables secure and sovereign data exchange with partners in ecosystems like Catena-X, enforcing the principles of the Trust and Governance pillars.
\end{itemize}

\section{Qualitative Evaluation}

Following the detailed presentation of CRACI’s components, we conducted both qualitative and quantitative evaluations of CRACI. In this section, we qualitatively assess its design by comparing it against the identified limitations of reviewed existing work discussed in Section II. This analysis demonstrates how CRACI’s architectural choices yield a more flexible, comprehensive, and operationally robust reference architecture for modern industrial systems.

\subsection{Overcoming Hierarchical Rigidity (compared to ISA-95 \& RAMI 4.0)}
A primary contribution of CRACI is its systematic resolution of the rigid, hierarchical structures that define both ISA-95 and RAMI 4.0. Where these models enforce a strict, top-down data flow that creates information silos, CRACI implements a data-centric \textit{hub-and-spoke} model. All semantically enriched data is channeled into the central \textit{Asset Representation Services} (the Digital Twin Hub), which acts as a single source of truth. Combined with an \textit{event-driven communication pattern} (pub/sub), this allows any authorized service to interact directly, eliminating the need for point-to-point integrations and enabling flexible, cross-layer communication. In order to overcome ISA-95's centralization of intelligence, CRACI distributes decision-making capabilities across the compute continuum. These services can be deployed from the low-latency edge (\textit{Operational Intelligence Services}) to central cloud (\textit{Strategic Intelligence Services}). Finally, the proposed CRACI addresses the analytical shortcomings of ISA-95 by separating the real-time operational path from the historical analytics path. This is enabled by a decoupled event bus, which allows the same data stream to be consumed in parallel by multiple specialized services. The Asset Representation Services subscribe to the stream to manage live, transactional state (an Online Transaction Processing (OLTP) workload). Simultaneously, a data sink service archives the stream into a long-term analytical store. This allows the \textit{Strategic Intelligence Services} to perform high-volume, historical analysis (an OLAP workload) without degrading the performance of the real-time system.

Furthermore, CRACI transforms the abstract concepts of RAMI 4.0 into concrete, operational capabilities. RAMI 4.0's conceptual \textit{Life Cycle \& Value Stream} axis is realized as CRACI's operational \textit{Lifecycle Management} pillar, which embeds DevOps and IaC practices. Similarly, where RAMI 4.0 only references external models for security and governance, CRACI embeds these as first-class \textit{Foundational Pillars}, ensuring that critical cross-cutting concerns are systematically enforced across every component in the system. The \textit{External and User-Facing Services}, specifically the \textit{Data Space Interface}, formalize external integration by supporting standards from ecosystems like the International Data Spaces Association (IDSA) and Gaia-X. This design provides a crucial foundation for addressing data sovereignty and compliance requirements, enabling secure and policy-driven collaboration in modern industrial networks.

\subsection{Providing a Complete and Technology-Agnostic Framework (compared to NOA \& FIWARE)}
CRACI extends the pragmatic but incomplete patterns of modern frameworks like NOA and FIWARE into a comprehensive, technology-agnostic architecture.

CRACI provides the missing M+O structure by decomposing it into clearly defined, distributed services. It explicitly defines \textit{Operational Intelligence Services} for low-latency analytics at the edge and \textit{Strategic Intelligence Services} for long-term optimization in the cloud, with their complex interactions managed by a dedicated \textit{Orchestration Services} component. Furthermore, CRACI extends NOA from a complementary pattern to a full-fledged system while keeping its core principle such as \textit{Verification of Request}.

CRACI generalizes the core concept of FIWARE's context broker into an architectural pattern, the \textit{Asset Representation Services} component, without prescribing the technology. This provides the same data decoupling benefits while allowing developers to implement the architecture with standard cloud-native tools (e.g., Kafka, Redis, custom APIs), thus avoiding lock-in. Crucially, CRACI also introduces native support for runtime orchestration, a key feature missing from FIWARE.

\subsection{Providing Implementation Guidance Beyond the Functional Models (compared to AIOTI Functional Model)}
The AIOTI model provides a comprehensive list of functional capabilities required for the Compute Continuum but remains architecturally abstract. Our architecture concretely implements these capabilities into specific system components. \textit{Data Management} and \textit{Artificial Intelligence} are realized through the \textit{Semantic Context Services} and \textit{Asset Representation Services} for data processing, and through \textit{Operational} and \textit{Strategic Intelligence Services} for edge and cloud analytics. \textit{Resource Management} and \textit{Networking} are addressed by the \textit{Core} and \textit{Edge Platform Services}, which manage compute, storage, and connectivity across heterogeneous environments. The \textit{Orchestration Services} component fulfills the orchestration function via workflow modeling and distributed process execution. Finally, instead of treating \textit{Security}, \textit{Privacy}, \textit{Trust}, and \textit{Monitoring} as separate blocks, our architecture elevates them to cross-cutting \textit{Foundational Pillars}—\textit{Trust}, \textit{Governance \& Policy}, and \textit{Observability}—ensuring these concerns are embedded throughout the system. By doing so, CRACI provides the necessary structure to turn AIOTI's functional view into an implementable system design.

\subsection{Embedding Cross-Cutting Concerns in a Detailed Architecture (compared to LASFA, SITAM, and 8C Architecture)}

CRACI can be viewed as the operational and engineering evolution of LASFA’s conceptual planning model. While LASFA offers an intuitive functional layout of smart factories, CRACI advances this by defining the operational principles and engineering patterns needed for a scalable implementation. It moves beyond component placement to mandate trust, observability, governance, and lifecycle automation as system-wide requirements, and refines LASFA’s decentralized “nested clouds” into a hub-and-spoke data architecture with a central \textit{Asset Representation Service}, ensuring a single source of truth and consistent intelligence services.

SITAM has identified several cross-cutting topics such as ``Service-Oriented Architecture (SOA) Governance'' and ``Data Quality.'' However, SITAM presents these as overarching topics to be considered across its functional layers. CRACI evolves SITAM by formalizing these cross-cutting concerns into four \textit{Foundational Pillars} and by updating its SOA blueprint for a fully cloud-native environment. Whereas SITAM outlines a data-driven factory based on an Enterprise Service Bus-centric integration model and centralized analytics, CRACI modernizes these concepts with a decoupled, event-driven communication bus, cloud-native runtime management, and distributed intelligence spanning edge and cloud.

CRACI extends the 8C Architecture by transforming its conceptual facets of ``Coalition'', ``Customer'', and ``Content'' into governed, scalable engineering patterns. Whereas the 8C model highlights the importance of horizontal and lifecycle integration, CRACI operationalizes these concepts through concrete mechanisms such as federated identity management, policy-driven data sharing, and automated lifecycle management. By elevating these facets into four \textit{Foundational Pillars}, CRACI moves beyond thematic cross-cutting concerns to enforce system-wide principles that ensure security, traceability, and agility across multi-stakeholder industrial ecosystems.

The summary of our qualitative evaluations is summarized in Table~\ref{tab:tick_comparison_balanced_styled}.

\begin{table*}[t]
\centering
\caption{Feature Comparison of Industry 4.0 Architectures}
\label{tab:tick_comparison_balanced_styled}
\small 
\definecolor{lightgray}{gray}{0.9}

\begin{tabular}{@{} l c c c c c c c c c @{}}
    \toprule
    & 
    \rotatebox{90}{ISA-95 \cite{scholten2007road}} & 
    \rotatebox{90}{RAMI 4.0 \cite{hankel2015reference}} & 
    \rotatebox{90}{NOA \cite{de2020namur, klettner2017namur}} & 
    \rotatebox{90}{FIWARE \cite{alonso2018industrial}} & 
    \rotatebox{90}{AIOTI Model \cite{open_continuum_eucloudedgeiot_task_force_2024_11656674}} & 
    \rotatebox{90}{LASFA \cite{resman2019new}} & 
    \rotatebox{90}{SITAM \cite{kassner2016stuttgart}} & 
    \rotatebox{90}{8C Architecture \cite{jiang2018improved}} & 
    \rotatebox{90}{\textbf{Proposed CRACI}} \\
    \textbf{Architectural Feature} & & & & & & & & & \\
    \midrule

    Formal Industry Standardization & 
    \checkmark & \checkmark & --- & --- & --- & --- & --- & --- & --- \\

    Industry-driven Operational Model &
    --- & --- & \checkmark & \checkmark & \checkmark & --- & --- & --- & --- \\

    Academic Proposal &
    --- & --- & --- & --- & --- & \checkmark & \checkmark & \checkmark & \checkmark \\

    \midrule
    Cloud-Native Principles & 
    --- & --- & --- & $\circ$ & $\circ$ & $\circ$ & $\circ$ & $\circ$ & \checkmark \\
    
    Decoupled / Event-Driven & 
    --- & \checkmark & $\circ$ & \checkmark & $\circ$ & $\circ$ & \checkmark & \checkmark & \checkmark \\
    
    Secure IT/OT Integration & 
    $\circ$ & $\circ$ & \checkmark & $\circ$ & --- & --- & $\circ$ & --- & \checkmark \\
    
    Cross-Domain/Ecosystem Integration & 
    --- & \checkmark & --- & $\circ$ & $\circ$ & $\circ$ & \checkmark & \checkmark & \checkmark \\
    
    Cross-Cutting Concerns & 
    --- & $\circ$ & --- & --- & \checkmark & --- & \checkmark & $\circ$ & \checkmark \\
    
    Brownfield Support & 
    \checkmark & $\circ$ & \checkmark & $\circ$ & --- & --- & $\circ$ & --- & \checkmark \\
    
    Data Sovereignty / Federated Support &
    --- & $\circ$ & --- & \checkmark & \checkmark & --- & --- & $\circ$ & \checkmark \\

    Human-Centric Design &
    --- & --- & --- & --- & $\circ$ & $\circ$ & \checkmark & $\circ$ & $\circ$ \\

    Concrete Implementation Blueprint & 
    --- & --- & $\circ$ & \checkmark & --- & \checkmark & \checkmark & $\circ$ & \checkmark \\

    \bottomrule
\end{tabular}

\vspace{1ex} 
\begin{flushleft}
\small \textbf{Legend:} \checkmark~Fully Addressed or a Key Strength; \quad $\circ$~Partially Addressed or Acknowledged; \quad ---~Not a Primary Focus or Not Addressed.
\end{flushleft}
\end{table*}

\section{Feasibility and Performance Validation}

\begin{figure}[htbp]
    \centering
    \includegraphics[width=\linewidth]{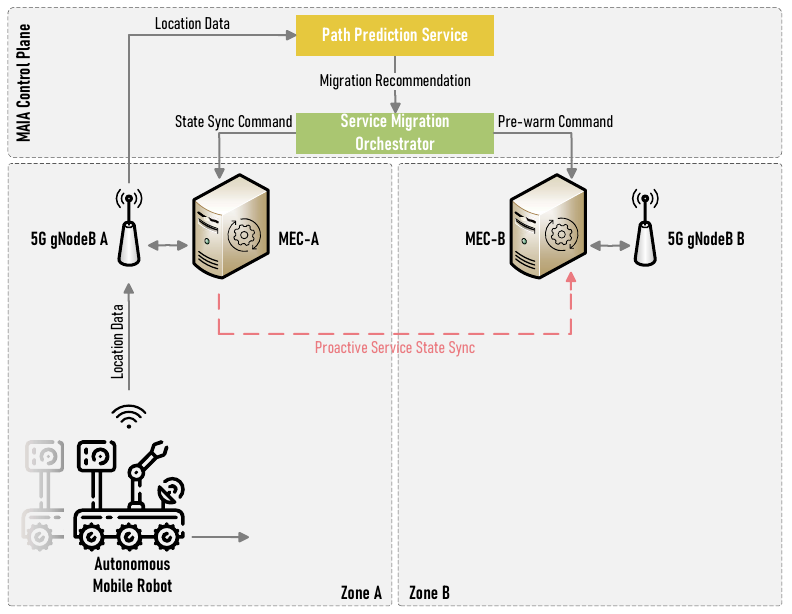}
    \caption{MAIA Use Case: Orchestration of Seamless Service Handover for Autonomous Mobile Robotics.}
    \label{fig:maia-usecase}
\end{figure}

The qualitative evaluationof CRACI demonstrates how the principles of existing models can be synthesized with established standards while embedding cloud-native design principles. To further assess the practicality of this design, we realized CRACI as a concrete implementation, \textit{MAIA} (Microservices-based Architecture for Industrial Data Analytics), and evaluated its performance. The results of this study have been published previously~\cite{dinh2019maia}. For detailed evaluation results, readers are referred to our previous work; this section focuses on discussing the results that directly validate the design of CRACI. MAIA was designed for a low-latency robotics use case as illustrated in Fig.\ref{fig:maia-usecase}, requiring proactive service migration for autonomous mobile robots (AMRs) moving between edge computing zones. Based on the location data collected from the AMRs, an edge-based predictive service anticipates their movement and proactively prepares the necessary services at the predicted destination. This enables seamless, low-latency service offloading for AMRs. This case study quantifies the performance and resource trade-offs inherent in a cloud-native industrial architecture and validates CRACI's design choices.

\subsection{Managing Resource Overhead through Architectural Governance}

A primary concern with microservice architectures is resource overhead. Our evaluation of MAIA confirms that while a naive \textit{lift-and-shift} containerization is inefficient, these overheads are a manageable trade-off when addressed through the disciplined practices defined by CRACI's Foundational Pillars.

Key findings from the MAIA evaluation show that by enforcing policies through the Lifecycle Management and Governance pillars, such as mandating lightweight base images and tuned JVM parameters, the resource overhead was drastically reduced (up to 80.74\% in artifact size and up to 78.79\% in memory consumption). Nonetheless, the results confirm an important reality that microservices inherently introduce overhead due to their distributed nature. Therefore, microservices represent a conscious architectural trade-off, exchanging some static resource efficiency for broader system-level benefits, namely modularity, resilience, and maintainability, which are essential in the context of complex industrial systems.

\subsection{Validating Performance and the Communication Bottleneck}

The end-to-end latency of the MAIA system was evaluated by simulating a fleet of up to 150 AMRs. Results show that latency remained below 50 ms across all test cases on a commodity server, confirming the architecture’s suitability for near–real-time industrial applications; for up to 100 robots, latency stayed under 20 ms. The analysis further revealed that scalability shifts the primary bottleneck from computation to coordination, as the share of time spent on inter-service communication grows from 67.62\% with a single robot to 86.97\% with 150 robots, while the internal processing time of each microservice remains nearly constant. This finding provides powerful validation for two of CRACI's core principles:

\begin{itemize}
    \item It underscores the absolute necessity of the Observability Pillar. Without deep instrumentation and tools like distributed tracing, such a communication-based bottleneck would be challenging to measure, especially in a distributed system across the continuum.
    \item It also provides a clear quantitative justification for CRACI's flexible deployment strategy. By placing time-sensitive Operational Intelligence Services at the edge, the architecture can minimize these critical network-based latencies, which is essential for low-latency industrial systems.
\end{itemize}

\section{Conclusion}

This work addressed the critical gap between traditional, hierarchical industrial architectures, such as ISA-95 and RAMI 4.0, and the demands of modern, cloud-native systems. We introduced CRACI, a comprehensive reference architecture for the compute continuum designed to be modular and scalable. Crucially, this work refrains from prescribing a specific technology stack, recognizing that while technologies evolve, the underlying architectural principles offer greater longevity. The primary novelty of CRACI lies in two areas: first, its data-centric, hub-and-spoke design, which systematically resolves the data silos of legacy models; and second, its formalization of four Foundational Pillars: Trust, Governance \& Policy, Observability, and Lifecycle Management, which elevate critical operational concerns to first-class architectural principles.

Through a detailed qualitative analysis, we demonstrated how CRACI's design systematically overcomes the limitations of existing standards, models, and proposals. Furthermore, we validated the architecture's practical feasibility through a quantitative analysis of MAIA, a real-world implementation. The evaluation showed that the inherent overheads of microservices (in artifact size and memory consumption) are manageable through the disciplined practices embodied by our Governance and Lifecycle Management pillars. The analysis also identified the communication fabric as the primary performance bottleneck at scale, underscoring the critical importance of the Observability pillar. 

Ultimately, CRACI should not be seen as a rigid, one-size-fits-all solution but rather as an architectural navigation system. It provides a robust and validated framework for engineering the next generation of secure, compliant, and agile industrial systems by offering a spectrum of decisions—from the latency requirements of edge services to the integration of brownfield assets. This approach is particularly well-suited for complex, heterogeneous environments like large-scale smart factories and integrated supply chains. Conversely, it may be over-engineered for simpler, static systems with hard real-time constraints, where monolithic embedded systems remain more appropriate. The contribution, therefore, lies in providing a structured approach to guide system architects toward more informed decision-making within the compute continuum.

The results of this study reveal key challenges that inform future research directions. First, inter-service communication was identified as the primary performance bottleneck, contributing up to 86.97\% of latency at scale. Addressing this requires the exploration of specialized messaging protocols and more efficient serialization formats, particularly suited for resource-constrained edge environments. Second, aligned with Industry 5.0 principles, there is a need to enhance human-centric intelligence and interaction within CRACI, shifting from basic data visualization toward collaborative human-machine environments that enable more intuitive and impactful operator engagement.

\bibliographystyle{IEEEtran}   
\bibliography{references}      

\end{document}